# On Quantifying Sentiments of Financial News

## *Are We Doing the Right Things?*

Gourab Nath
*Department of Data Science*
*Praxis Business School*
Bengaluru, India
gourabnath88@gmail.com

Arav Sood
*Department of Data Science*
*Praxis Business School*
Bengaluru, India
aravsood10@gmail.com

Aanchal Khanna
*Department of Data Science*
*Praxis Business School*
Bengaluru, India
aanchal.khanna@praxis.ac.in

Savi Wilson
*Department of Data Science*
*Praxis Business School*
Bengaluru, India
savi.wilson@praxis.ac.in

Sree Kavya Durbaka
*Department of Data Science*
*Praxis Business School*
Bengaluru, India
sreekavya1412@gmail.com

Karan Manot
*Department of Data Science*
*Praxis Business School*
Bengaluru, India
karan.manot@praxis.ac.in

*Abstract*—Typical investors start off the day by going through the daily news to get an intuition about the performance of the market. The speculations based on the tone of the news ultimately shape their responses towards the market. Today, computers are being trained to compute the news sentiment so that it can be used as a variable to predict stock market movements and returns. Some researchers have even developed news-based market indices to forecast stock market returns. Majority of the research in the field of news sentiment analysis has focussed on using libraries like Vader, Loughran-McDonald (LM), Harvard IV and Pattern. However, are the popular approaches for measuring financial news sentiment really approaching the problem of sentiment analysis correctly? Our experiments suggest that measuring sentiments using these libraries, especially for financial news, fails to depict the true picture and hence may not be very reliable. Therefore, the question remains: What is the most effective and accurate approach to measure financial news sentiment? Our paper explores these questions and attempts to answer them through SENTInews: a one-of-its-kind financial news sentiment analyzer customized to the Indian context.

I. INTRODUCTION

Over the past decade, India, as an economic nation has grown exponentially. The stock market has boomed for the greater good of two main factions - the investors who see it as a primary source of income and the common people who want to generate a passive income. This stock market rise has its own shortcomings in the form of volatility and adding to this is the uncertainty caused by the sentimental significance of the news articles about various companies listed in the stock market. Since the covid-19 pandemic, it is evident now (if not before) that historical indicators alone are not enough in gauging the performance of the stock market. Moreover, these indicators are not very intuitive and hence very hard to understand for a common man. For example, let's say a reader reads a news article with the headline "Nykaa hits a fresh low, but this brokerage sees 57% potential upside in stock". It's not easy for a reader, especially if he or she is a common man, to understand the sentiment of this statement. Moreover, the sentiment of a reader may also be associated with personal bias towards the company. This makes him second-guess his own opinion, often resulting in an incorrect and delayed decision.

Furthermore, a news article is divided broadly into three parts i.e., its headline, a synopsis and finally the full body of the news article. Sometimes a news headline, written primarily to attract a reader's attention, might not be an accurate reflection of the news. While it is important to read the news, lengthier news may make it difficult for a general reader to retain the context especially with the presence of contradictory statements.

Due to the complex and subjective nature of sentiment analysis, market sentiment is difficult to quantify and define. As a result, various techniques for computing sentiments can disagree with each other, sometimes to a great extent. In this project, we offer several examples from the results of our well-designed experiments, of how conventional approaches may fail to capture the sentiment and polarity of a news article, especially in the case of financial news. Apart from ranking the performances of various conventional approaches, we suggest a robust way of computing news polarity for financial news through our financial news sentiment analyzer - SENTInews.

II. LITERATURE REVIEW

Existing research on sentiment analysis can be broadly classified into two approaches, the former being the most widely applied method for textual analysis:

*A. Rule based or lexical text sentiment analysis*

Rule-based methodologies are based on rigid and static semantic rules. These involve matching words present in a news article against a predefined vocabulary of positive and negative words denoting sentiments. Depending on the frequency of positive vs negative vocabulary words in the chosen text, sentiment scores are calculated and assigned. Additionally, to improve performance of these rule-based libraries on a diverse corpus of texts, adjustments have been made such that the predefined vocabulary varies according to the domain of analysis. For instance, prevailing sentiment analysis libraries such as VADER and Pattern are generally used for social media texts [2,15] whereas libraries such as Loughran-McDonald, Harvard General Inquirer are used for texts pertaining to the financial domain [3,14].

Rule-based approaches pose challenges due to their limited vocabularies and their inability to capture context, especially in the case of financial news. They have also been found to routinely underperform in capturing the nuances in the news articles and in predicting the sentiment correctly [2,6]. To overcome the above mentioned challenges, many alternate frameworks have proposed modifications to the existing libraries by way of enhancing lexicons and introducing contextual components [2,3,5]. Our paper also presents two such enhanced frameworks of rule-based libraries: 1) modified VADER and, 2) modified LM.

*B. Machine Learning based text sentiment analysis*

Various machine learning approaches have sought to overcome the problem of non-contextualization in sentiment analysis. [7,8,10] use deep neural networks like RNN and BERT to produce dynamic, contextualised embeddings, which are essentially encoded vector representations of the news articles that capture the complexity of context, word positioning, word order, etc. This is in contrast to techniques like Bag of Words and TF-IDF [2,9], Doc2Vec which are simpler but they yield static embeddings devoid of contextual information. Both types of word embeddings can be used as inputs into a myriad of machine learning models like Naive Bayes, Logistic Regression, LSTM, XGBoost [4, 10]. In our experiments, we compare the performances of various ML models using different static vs contextualised embedding approaches. Finally, the chosen framework for SENTInews relies on the superior performance of dynamic embeddings in a customised deep learning model setup.

While DL models tend to provide better performance, these methods are time-consuming and computationally expensive, not to mention their requirement of large labelled datasets for training. [1] Retrieving such large, pre-labelled news datasets, especially relevant to a particular region (here, India), is not easy. Due to this, researchers often have to laboriously label hundreds to thousands of news articles [7]. Moreover, in order for manual labelling to yield meaningful labels, it is imperative that thorough sampling and grading frameworks are established. In our research, we carefully consider these parameters, and set up a data collection pipeline to scrape 1.4 million news articles, followed by a sampling methodology to create a robust, labelled dataset. Our framework encapsulates a custom tokenizer, a deep learning approach to contextualise the embeddings followed by the downstream task of sentiment analysis. This encapsulation forms the basis of our paper's major contributions to existing literature.

III. METHODOLOGY

*A. Data Collection and Preparation*

*1) Data Collection*

An automated data scraping framework based on parallel processing is designed to retrieve news articles from *The Economics Times*. The data scraping framework uses a web scraper to scrape the news articles from the website. The web scraper is created using the beautifulsoup [18] library of Python and the data extraction framework is created using concurrent futures [19] and parallel jobs packages for multiprocessing. Our framework helped us to collect all the daily news articles from the year 2008 to the year 2022 for a total of 178 months. The total number of news articles accumulated is 1,397,114. Each instance in our data represents a news article and contains features like publish date & time, update date (if any), headline, synopsis, sector/genre (e.g. entertainment, national, economy & banking, international, etc.) and the full text of the news article.

*2) Data Preprocessing*

The news articles scraped from the web are further sent to an automated data pre-processing pipeline to perform the necessary functions before the final ready-to-use news data is returned. The news articles with missing synopsis and full text are dropped. The regex function is used to remove superfluous characters such as numbers, punctuations and special notations. NLTK stopwords are removed, barring words like up, above, etc. which are relevant for stock market sentiment analysis. Texts are converted to lowercase to ensure standardisation. A column is created by merging the headline and synopsis together for each news. The 'full text' of any news articles that are shorter than 20 characters is replaced by NaN. A concise list of news sectors is created. The top four most frequently occurring sectors remain with their original names whereas all other sectors are clubbed in the 'other' category. This further helps us to segment the dataset into financial or non-financial news. Since the pre-processing stage is computationally intensive, the entire dataset is split into multiple datasets at a five-year gap to support faster processing.

*3) Sampling and Text Annotation*

To make any statistical inferences about the efficacy of various sentiment analysis approaches, we need to manually label the data to compare the performances of these approaches. To proceed further from there, we group the news articles into two major categories - financial news and non-financial news. Since manually labelling each news is an intensive task we have kept our scope smaller by reducing the population of interest to the set of news articles that are published in the first quarter of the year 2022 (i.e. total of 36,851 articles) considering their recency at the time of research and lower influence due to the Covid-19 pandemic. Within this population of interest, the relative frequency of the genres/sectors is as follows: other - 46.3%, stocks - 24.5%, politics & India - 16.3%, international - 8.1%, economy & banking - 4.7%. Next, the population of interest is further segmented into financial and non-financial news. Financial news comprises those articles with the genre tags 'economy&banking' and 'Stocks' (total of 10,758 articles), whereas the remaining articles are categorised under non-financial news (total of 26,093 articles). A sample size calculator calculates the sample size by setting the confidence level as 95%, and the margin of error as ± 5%. The calculated sample size is 385 and is rounded off to 400 for each segment. Since the proportions of genres within financial and non-financial news are different, we use stratified random sampling to generate samples of 400 each from both the segments. For instance, the 'stocks' genre is dominant with 84% frequency in the financial news segment, so a stratified sample of 336 articles is drawn from this genre, whereas a sample of 64 articles is drawn from the 'economy&banking' genre.

Three proficient readers within the team with formal training in Economics, and with fair knowledge of trading in the stock market are assigned to the text labelling tasks independently. Each human tagger is expected to read all the 800 sampled news articles and rate them discreetly on a scale that ranges between -2 and +2. The values -2 and -1 represent very negative and moderately negative respectively whereas the values 1 and 2 represent moderately positive and very positive. This scale doesn't include the value 0 and hence there is no option to mark a news as neutral. This is done based on the assumption that every piece of news carries a sentiment. The final score for each news article is an aggregate of the scores given by the three independent reviewers. A column with the labels 'positive' and 'negative' is also created based on whether the average news score is more or less than 0 respectively. Active steps are taken to preserve the integrity of the sampling process. During this process, the human reviewers are excluded from any research works that are related to lexicon-based sentiment analysis to eliminate biases. A unique encryption ID is generated to mask the date of the news published.

This framework poses its own challenges due to the paucity of time and labour. Additionally, there is high subjectivity involved in labelling articles when full text is considered, as a single article could have conflicting news components. However, as mentioned above, in order to make any statistical inferences regarding the application of existing sentiment analysis techniques, it is considered to be the right way to approach the problem.

*B. Lexicon-based sentiment analysis*

The use of rule-based or lexicon-based approaches like Vader [15] and Pattern [6] is a very common and popular choice especially when it comes to analysing the news sentiment. For analysis of financial news, libraries like Harvard-IV and Loughran-McDonald [14] that include some special words having financial and economic importance are proposed. In this work, we have proposed two additional dictionaries - modified LM and modified Vader that are formed by combining the financial and non-financial lexicons.

*1) General lexicon for sentiment analysis*

VADER (Valence Aware Dictionary and Sentiment Reasoner) is a rule-based sentiment analyzer that is domain agnostic as its wordlists are derived from social media or microblogging. It has approximately 7500 words along with lexical features like emoticons (e.g. :D and :P ), sentiment-related acronyms (e.g. LOL and ROFL), and common slang words with sentiment value (e.g., Nah and meh) [15,17].VADER incorporates the impact of each sub-text on the perceived intensity of a sentiment in the sentence-level text known as Heuristics. Punctuation, capitalisation, degree modifiers, polarity shift due to conjunction, and polarity negation are the five rules used in VADER[15,17]. It computes the compound score by aggregating the valence scores. The word list has scores ranging from + 4 to -4 and subsequently normalised scores in VADER range from -1 to +1 where -1 represents the most extreme negative sentiment and +1 represents the most extreme positive sentiment [15,17].

Other dictionaries, Pattern and Harvard IV are popular sentiment analysis tools. However, their documentation of wordlists is not accessible for in-depth analysis. Pattern generates two levels of output i) Polarity scores, which measure positive vs. negative, on the scale of -1.0 and +1.0 and ii) subjectivity, which measures objective vs. subjective as a value between 0.0 and 1.0. Harvard is a general-purpose dictionary with two large valence categories Positive & Negative, derived from applications in psychology and sociology. Loughran & McDonald report that almost 75% of the words contained in Harvard IV negative wordlist do not convey a negative sentiment when used in a financial context. [14]

*2) Finance-related lexicon for sentiment analysis*

Loughran-Mcdonald (LM) Dictionary is one of the most popular financial domain-based dictionaries created by Tim Loughran and Bill Mcdonald [14]. Approximately 4,000 lists of words were considered from the U.S. Security and Exchange Commission portal over the period of 1994 to 2008. The wordlists are further classified into 6 categories, viz. negative, positive, litigious, uncertain, strong modal, and weak modal. The scoring mechanism of LM assigns a +1 score to words classified as 'positive' and a -1 score to words classified as 'negative'. Categories other than positive and negative are not given any weightage in sentiment analysis [14]. LM dictionary-based sentiment analyzer generally gives three levels of output i). Word count of the positive and negative words, polarity and subjectivity scores [14].

Polarity score is calculated as a ratio of net positive divided by total positive and negative words in a sentence as in:

$$\text{Polarity} = (Pos-Neg)/(Pos+Neg) \qquad (1)$$

Subjectivity score is calculated as the ratio of total positive and negative words to the total number of words in a sentence as in:

$$\text{Subjectivity} = (Pos+Neg)/N \qquad (2)$$

*3) Modified lexicons for sentiment analysis*

Due to the subjective nature of these lexicons, it is observed that there are certain words which are included in one lexicon that do not exist in the other lexicon. For example, there aren't words in the VADER lexicon that can analyse the sentiments of the financial news well. On the other hand, LM doesn't contain enough words to capture the general sentiments of an article. In this research, we experimented by creating two modified lexicons, LM-in-VADER and VADER-in-LM. As the names suggest, LM-in-VADER is created by adding the words from the LM lexicon to the existing VADER lexicon while the VADER-in-LM lexicon is created by adding the VADER lexicon to the existing LM lexicon.

A comparative analysis of VADER and LM lexicons provided the ground for modifying these libraries. For words present exclusively in VADER, we maintained the original score of VADER. It ranges from -4 to +4 where a negative value up to -4 denotes extreme negative sentiment and positive value up to +4 for extreme positive sentiment, and 0 for neutral sentiment [5,15]. For the common words between LM and VADER, as well as from words exclusively from the LM lexicon, we maintained the valence scores of original LM which range from -1 to +1. [14]. This however did not serve the purpose of enhancing the performance of the VADER lexicon because VADER has a large corpus of words compared to LM and the words in VADER have higher valence scores to begin with.

For VADER-in-LM, a similar approach is applied as above, and the weightage originally assigned by LM [14] is maintained for each word including the common words between LM and VADER. We did not see a significant improvement in the performance as there is an inherent limitation of the size of the wordlists and the neutralising tendency in providing sentiment scores. Specifically in the financial news context, an article can talk about how the stock prices of XYZ company plummeted because of some negative reason in a quarter and increased now. Lexicon-based sentiment analysis fails to capture the context of such sentences while aggregating the polarities of individual words.

*C. Model-based sentiment analysis:*

For any text to be considered as input to a machine learning model, it needs to be broken down into smaller sub-texts (or tokens) and converted to a vector (which is a numerical representation of these tokens) that can eventually be fed into a model. There are diverse methods of converting longer texts into tokens, and the superiority of each method determines the quality of information retained in the sub-texts after tokenization. In this section, our model based experiments are categorised on the basis of four different tokenization techniques.

*1) Model based on word tokenizers*

Word tokenizers are the most basic tokenizers as they create tokens by splitting the text into sub-texts (here, words) based on white space.

TF-IDF is one of the most popular and simplest vectorization techniques that utilises word tokenization [9,10]. The advantage of this method is that it is easy to interpret and use, however its disadvantage is that it creates sparse vectors which are devoid of contextual information. Additionally, as each unique word appearing in the text is a feature, the column to row ratio is infinitely high in the data, thus, compelling us to deal with the problem of dimensionality. After converting each news as TF-IDF vectors, supervised learning models like naive bayes and decision tree, bagging, random forest, etc. are used for sentiment classification. The logistic regression model is not used considering the high dimensionality of the data.

*2) Model based on character tokenizers*

To try another way to deal with the problem of contextualization in the above-mentioned approaches, we explore a pre-trained ELMo model [7] to create embeddings (vector representations) that can be used as model input. This technique uses character level tokenization, which is better equipped to handle out-of-vocabulary (OOV) words [22]. Within the ELMo architecture, the tokens are first fed into an embedding layer and then passed through a Bi-LSTM layer to retain semantic and contextual information. However, in our experiment, we use ELMo as a vectorizer and not as a sentiment classifier. The contextualised embeddings obtained from the ELMo model are passed through logistic regression and random forest classifiers to carry out the downstream tasks of sentiment classification. Since this method uses character level tokens, its major disadvantage lies in the computational complexity owing to the length of the vector created for each text. Also, previous research shows that ELMo underperforms on longer texts [16]. To combat the issue of vector size, there do exist variations of the pre-trained ELMo models that can be used, however, they are extremely computationally expensive. Another disadvantage of this pre-trained model is that it has been trained on a specific, limited corpus of text whose domain we are unaware of.

*3) Models based on sub-word (word-piece tokenizer:*

The BERT model's advantages are that it uses a form of sub-word tokenization - that is WordPiece tokenization - to preserve semantic/contextual richness of the input text. The model also has a more superior way of dealing with OOV words [22] as it doesn't unnecessarily increase the vector lengths when it creates WordPiece tokens. The BERT model is designed using transformers that are built on the encoder-decoder architecture [10,23]. It can also deal with supervised learning problems as the last layer of the model can be altered depending on the regression or classification problem at hand. As is the case in ELMo, BERT models are plagued with issues of time and computational complexity [7]. The pre-trained model variation we use in this experiment restricts input vector lengths to 512, and is also trained on an unknown corpus. These pitfalls inhibit the usage of larger texts such as full text, and pose the risk of loss of contextual information.

*4) SENTInews: The Proposed Model Framework*

*"Fig 1"* Our proposed model framework attempts to provide a solution by addressing three dominant issues faced in our model based approaches. These issues are, 1) dealing with OOV words , 2) retaining as much contextual and semantic information as possible, 3) reducing the computational complexity. We begin by training a customised, special kind of sub-word tokenizer called WordPiece tokenizer [23] on 371,058 financial news articles. This is in contrast to the pre-trained BERT tokenizer which is trained on an unknown domain/dataset. WordPiece tokenization creates new phrases by learning merge rules for characters based on maximum likelihood, and not on the most frequent occurrence of sub-words or phrases. For example, WordPiece will choose "de" as a merged sub-word only if the probability of "de" divided by "d", "e" would be more than any other merged sub-word once added to the vocabulary. Once "de" is added to the vocabulary, it can single-handedly account for words like decline, debt, debenture, debit etc. without any need for having separate tokens for these words. This helps in creating a diverse vocabulary with meaningful tokens without unnecessarily increasing the vocabulary size. As a result, it reduces the memory and computational requirements of training a tokenizer.

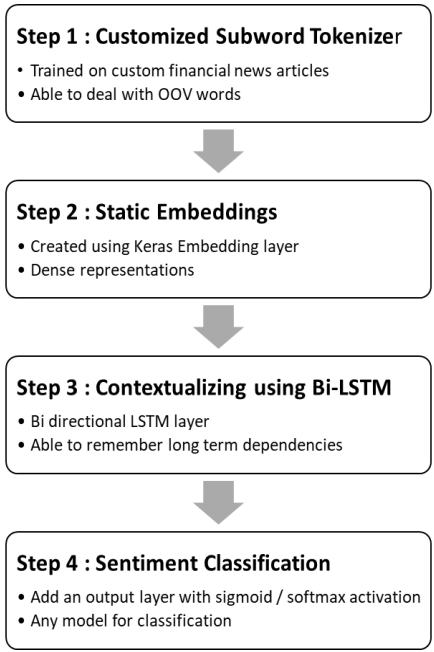

Fig. 1.  SENTInews – the proposed model framework

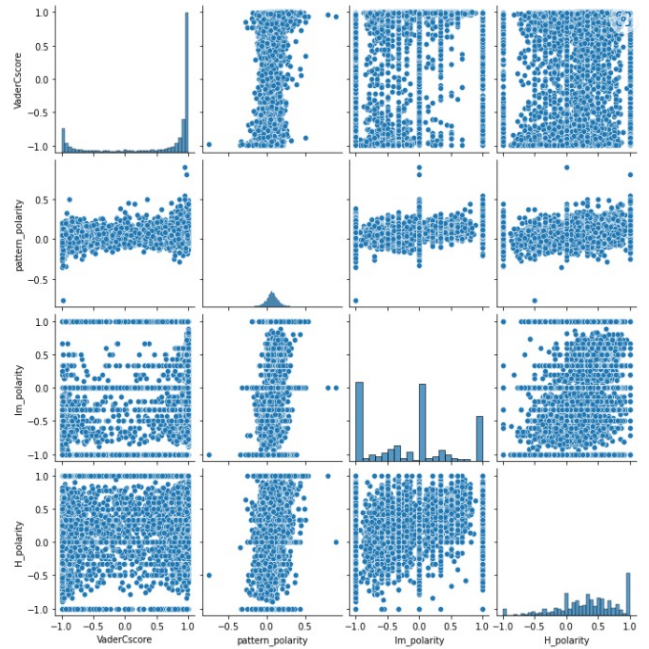

Fig. 2.  Pairwise Correlation

Next, the resulting tokens from the custom tokenizer are converted into a sequence, also referred to as a dense embedding, using a keras embedding layer. This static embedding is contextualised using a Bi-LSTM architecture which not only helps in learning the patterns present within the text but also retaining long-term contextual information. Lastly, a dense keras output layer consisting of a sigmoid activation function is added to complete the downstream task of binary sentiment classification (positive/negative).

IV. RESULTS

The major findings of this paper are categorised into three sections: 1) Comparison of rule-based approaches, 2) Comparison of ML approaches, 3) SENTInews.

A. *Comparative analysis of lexicon-based sentiment analysis*

*1) Correlation analysis of lexicon-based approaches*

For this study, we randomly chose the year 2021. Six months were selected at random, and subsequently, five days were chosen at random from each of these six months, resulting in a dataset comprising 30 days. Articles corresponding to the selected dates were retrieved and processed using specialised functions available within various libraries. The aggregated data from the dictionaries was then formatted into a data frame.

*"Fig. 2"* We then plotted a correlation graph for Harvard IV, LM, Pattern and Vader. VaderCscore is the compound score given by Vader. Pattern_polarity, lm_polarity and H_polarity are the polarity scores given by Pattern, LM, and Harvard IV respectively. We found that Vader has the highest correlation with all other libraries. A high categorization of extreme positives and negatives is observed in Vader. Pattern displayed normally distributed graphs. LM and Harvard IV are positively correlated. Based on this association, Harvard IV was eliminated from further analysis.

*2) Comparative study of popular lexicon libraries*

Among the two dictionaries, Vader & LM, 930 words were found in common as shown *"Fig. 3"*. Among total unique negative words (6515), 662 are common between both libraries *"Fig. 4"*. On the other hand, only 179 unique positive words are common *"Fig. 5"*. A total of 20 negative words from LM are considered positive in Vader and only 1 word is considered positive in LM and negative in Vader. The polarity mismatch is very less between these libraries.

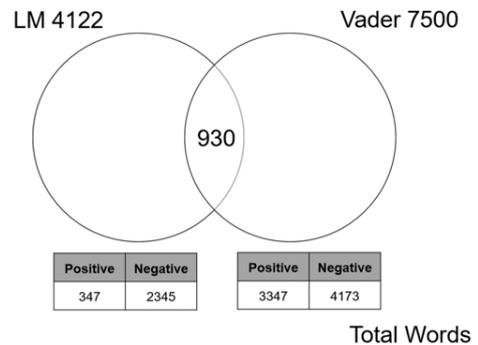

Fig. 3.  Total Words in LM and Vader

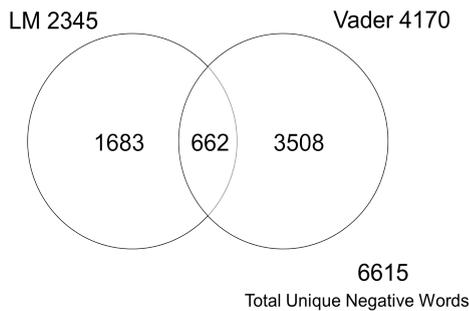

Fig. 4. Unique Negative Words in LM, Vader

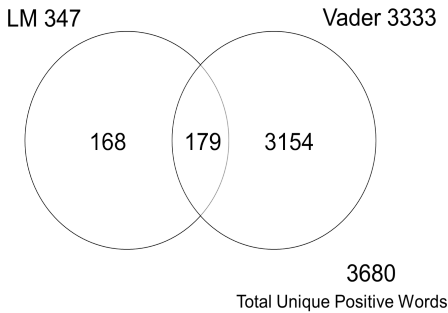

Fig. 5. Unique Positive Words in LM, Vader

*3) Comparison of lexicon-based sentiment analysis*

*"Table I"* Following the implementation of modified versions of Vader and LM, we assess the performance of both the existing and modified dictionaries on labelled data. The modifications applied to both VADER and LM result in improved accuracy, showcasing the potential of these modifications. For financial news, the modified version of LM, i.e. VADER-in-LM demonstrates improved accuracy compared to the original LM for both full text and headline+synopsis. Surprisingly, we also observe substantial improvement of VADER-in-LM in evaluating the sentiment of full text of non-financial news (70.7%) compared to original LM (58%). LM-in-VADER does not show significant changes to performance. The sentiment analysis of full texts of non-financial news exhibited the highest accuracy (76.5%) with the original VADER. Overall, it is evident that both VADER and Pattern fall short in exhibiting consistent and competitive performance in sentiment analysis of financial news content. Considering the accuracy scores, even LM and its modified version demonstrate scope for improvement in sentiment analysis of financial news

TABLE I. RESULTS OF THE LEXICON-BASED APPROACH

| Name | Accuracy Scores | | | | | |
|---|---|---|---|---|---|---|
| | *A* | *B* | *C* | *D* | *E* | *F* |
| Vader | 51.7 | 58.5 | 64.0 | 76.5 | 58.5 | 51.7 |
| LM | 35.2 | 50.2 | 42.5 | 58.0 | 50.2 | 35.2 |
| Pattern | 44.0 | 56.5 | 49.5 | 64.5 | 56.5 | 44.0 |
| LM-in-VADER | 52.0 | 59.7 | 68.2 | 74.7 | 59.7 | 52.0 |
| VADER-in-LM | 45.7 | 58.0 | 57.7 | 70.7 | 58.0 | 45.7 |

A - Sentiment of Headline+Synopsis of Financial news
B - Sentiment of Full text of Financial News
C - Sentiment of Headline+Synopsis of Non Financial news
D - Sentiment of Full text of Non Financial News
E - Sentiment of Full text of All news
F - Sentiment of Headline+Synopsis of All news

*B. Comparison of ML approaches*

*1) Models based on word tokenizers*

TF-IDF is implemented on the features 'full text' and 'headline + synopsis' of 400 financial news articles, which yield sparse vector representations and dense vector representations respectively. The experiment is repeated for 400 sampled non-financial news. Sparse vectors are fed into three types of models: decision tree, random forest as well as naive bayes classifier models.

*"Table II"* The chosen models with TF-IDF vectorization technique did not yield significant results for non-financial news. *"Table III"* for financial news, it is observed that the decision tree family classifiers are susceptible to high variance as the data size is very small. Bagging model is implemented to reduce the variance in the decision trees, thereby pushing up the test accuracy to 73% for financial news full-text and 74% for financial news headline + synopsis.

*2) Models based on character tokenizers*

*"Table IV"* ELMo uses character level tokenization to retain semantic and contextual information. For this experiment, we use the 'headline + synopsis' and 'full text' features of the full combined sample of financial and non-financial news to obtain ELMo embeddings. The embeddings are implemented in logistic regression and random forest models for the combined sample of financial+non-financial news and then later on, for financial and non-financial news separately.

TABLE II. RESULTS OF ML MODELS WITH TF-IDF EMBEDDING (NON-FINANCIAL NEWS SAMPLE)

| Model | max_depth | Accuracy Score | | | |
|---|---|---|---|---|---|
| | | *Full Text* | | *Headline+Synopsis* | |
| | | *train* | *test* | *train* | *test* |
| Decision Tree | 3 | 0.77 | 0.68 | 0.68 | 0.60 |
| Random Forest | 5 | 0.74 | 0.54 | 0.63 | 0.51 |
| Naive Bayes | None | - | 0.53 | - | 0.53 |

TABLE III. RESULTS OF ML MODELS WITH TF-IDF EMBEDDING (FINANCIAL NEWS SAMPLE)

| Model | Accuracy Score | | | | | |
|---|---|---|---|---|---|---|
| | Full Text | | | Headline+Synopsis | | |
| | max_depth | train | test | max_depth | train | test |
| Decision Tree | 4 | 0.87 | 0.65 | 5 | 0.78 | 0.65 |
| Random Forest | 5 | 0.87 | 0.61 | 5 | 0.77 | 0.61 |
| Bagging | 5 | 0.93 | 0.73 | 6 | 0.88 | 0.74 |
| Naive Bayes | None | - | 0.60 | None | - | 0.68 |

"Table V,VI" The results we observe in ELMo embeddings+Logistic Regression are consistently better across various sample choices where the model input is the headline + synopsis feature. In comparison, the lower performance on full text is consistent with the limitations of ELMo embeddings highlighted in the methodology section. We attain the highest accuracy at 84% for the full sample using headline + synopsis. At this point, it becomes clear that contextualization using Bi-LSTM architecture therefore provides an effective solution in exploring the best approach towards sentiment analysis of financial news.

3) *Models based on sub-word tokenizers*

The first BERT trial was conducted on the full text feature of the full combined sample of financial and non-financial news due to similar reasons outlined in the previous section. In case of full sample, we achieved a test accuracy of 0.51 by running different (small) batch sizes and varying learning rates. The model was computationally complex and time-intensive.

TABLE IV. RESULTS OF ML MODELS WITH ELMO EMBEDDINGS (FULL SAMPLE)

| Model | Accuracy Scores | | | |
|---|---|---|---|---|
| | Full sample | | | |
| | Training Accuracy (Full text) | Test Accuracy (Full text) | Training Accuracy (Headline+ Synopsis) | Test Accuracy (Headline + Synopsis) |
| Logistic Regression | - | 0.76 | 0.94 | 0.84 |
| Random Forest | - | 0.69 | 1 | 0.75 |

TABLE V. RESULTS OF ML MODELS WITH ELMO EMBEDDINGS (FINANCIAL NEWS SAMPLE)

| Model | Accuracy Scores | | | |
|---|---|---|---|---|
| | Financial news sample | | | |
| | Training Accuracy (Full text) | Test Accuracy (Full text) | Training Accuracy (Headline+ Synopsis) | Test Accuracy (Headline + Synopsis) |
| Logistic Regression | 0.79 | 0.60 | 0.93 | 0.79 |
| Random Forest | 1 | 0.65 | 1 | 0.74 |

TABLE VI. RESULTS OF ML MODELS WITH ELMO EMBEDDINGS (NON-FINANCIAL NEWS SAMPLE)

| Model | Accuracy Scores | | | |
|---|---|---|---|---|
| | Non-financial news sample | | | |
| | Training Accuracy (Full text) | Test Accuracy (Full text) | Training Accuracy (Headline+ Synopsis) | Test Accuracy (Headline + Synopsis) |
| Logistic Regression | 0.88 | 0.74 | 0.97 | 0.75 |
| Random Forest | 1 | 0.76 | 1 | 0.68 |

Since the training dataset is small, this neural network seems to be one of the lowest performing models seen in our research so far. There could be multiple reasons for the underperformance. For one, the BERT tokenizer has a limited vocabulary and only outputs 512 tokens. Many instances of our full text feature are very lengthy, and the required semantic information may not be captured with these limited tokens. Additionally, the pre-trained BERT model operates on a tokenizer which has been trained on a general corpus, not necessarily relevant to news articles. Furthermore, we also considered training a BERT model from scratch, however, this experiment was paused midway as training took multiple hours and required greater computational power. To counter the possible reasons for underperformance, we would like to replicate this experiment on shorter input text, i.e., our headline + synopsis feature, in the future.

4) *SENTInews*

Time and again in our previous experiments, we observe the need to develop a deep learning model that balances complexity (time and computational), contextualization (retention of context over longer sentences and learning semantic patterns) and correctness (classifying sentiment with high accuracy scores). We explore the performance of SENTInews' framework on full text and headline + synopsis features of the financial news. We do not implement this on non-financial news as the custom tokenizer is trained on financial news corpus only.

TABLE VII. RESULTS OF SENTINEWS MODEL (CUSTOM TOKENIZER + KERAS EMBEDDINGS + BI-LSTM) ON FINANCIAL NEWS

| SENTInews Model | Accuracy Scores | | | |
|---|---|---|---|---|
| | Training Accuracy (Full text) | Test Accuracy (Full text) | Training Accuracy (Headline+ Synopsis) | Test Accuracy (Headline+ Synopsis) |
| Trial 1 | 0.86 | 0.55 | 0.98 | 0.73 |

*"Table VII"* We achieve an accuracy of 73% for sentiment classification for the headlines + synopsis feature of financial news through SENTInews analyzer. Overall, for financial news, rule-based libraries yield accuracy scores ranging between 50-60%, whereas comparable word tokenizer models provide an accuracy of about 73% on full text (bagging model) and 79% on headline + synopsis (ELMo embeddings + logistic regression).

Irrespective of tokenizers, neural networks outperform rule-based libraries for sentiment classification of financial news. However, amongst all the neural networks, our proposed framework clearly balances complexity, contextualization and correctness. It also emphasises the importance of an efficient way to build vocabulary for the purpose of tokenization for sentiment analysis. Therefore, these results provide an indication to enhance our manually labelled dataset such that we obtain an increased size of training data. Although the proposed model gives below average results for full text, this can be attributed to the simplicity of the model and can be improved by increasing the size of input layer, adding another bidirectional layer i.e. increasing model complexity gradually. As neural networks can perform better on more data, we see potential in creating a more robust version of our proposed model which is trained on a larger dataset.

V. CONCLUSION

We conducted exhaustive research into existing sentiment analysis methods, categorising them into rule-based and machine learning-based approaches. Rule-based methods, although widely used, were insufficient in analysing financial news due to their limited vocabularies and inability to grasp complex context. On the other hand, machine learning-based approaches, particularly deep learning models, showed potential in understanding sentiment in context. After comparing performances of various sentiment analysis approaches, we observed that it was crucial to introduce a news sentiment analyzer which aimed at quantifying financial sentiment objectively while eliminating personal biases and maintaining context. Hence comes in SENTInews, which demonstrates one of the highest accuracy scores amongst our model-based experiments on financial news, notably outperforming popular sentiment analysis libraries.

The implications of our findings are extraordinary. The proposed methodology holds the potential to redefine sentiment analysis in financial news, offering more accurate and reliable insights. If deployed, a more robust and refined version of this tool can assist investors to make better informed and timely decisions, potentially optimising their investment strategies. In summary, Senti5 represents a significant stride in advancing sentiment analysis.